\documentclass[english,fleqn]{article-hermes}
\usepackage{graphicx}
\usepackage{journalabbrev}

\newcommand{\microns}{\ensuremath{\mbox{$\mu$m}}}

\proceedings{Visions in InfraRed Astronomy. Volume 5, n$^\circ$ 3-4/2005}{1}
\title[Tomorrow optical interferometry]{Tomorrow optical
  interferometry: astrophysical prospects and instrumental issues} 
\author{Fabien Malbet}

\address{%
Laboratoire d'Astrophysique de Grenoble\\
CNRS / Université Joseph Fourier (Grenoble 1)\\
BP 53, F-38041 Grenoble cedex 9, France\\[3pt] 
Fabien.Malbet@obs.ujf-grenoble.fr}

\resume{L'interférométrie a permis d'apporter de nouvelles contraintes
  en astronomie optique dans les dernières années. Dans ce domaine,
  l'ouverture d'interféromètres à très grandes ouvertures comme ceux du
  Very Large Telescope européen ou du télescope Keck américain a
  permis d'effectuer un énorme bond en avant. Penser le développement
  futur est à la fois facile --la plupart des spécialistes savent dans
  quelles directions se diriger pour développer l'interférométrie-- et
  difficile à cause de la complexité croissante de cette technique. Je
  présenterai un court état des lieux de l'interférométrie
  aujourd'hui. Ensuite je détaillerai les perspectives astrophysiques
  envisageables. Finalement je m'attacherai à évoquer quelques points
  spécifiques à l'instrumentation qui sont décisifs
  pour l'avenir de l'interférométrie.}
\abstract{Interferometry has brought many new constraints in optical
  astronomy in the recent years. A major leap in this field is the
  opening of large interferometric facilities like the Very Large
  Telescope Interferometer and the Keck Interferometer to the
  astronomical community. Planning for the future is both easy --most
  specialists know in which directions to develop interferometry-- and
  difficult because of the increasing complexity of the technique. I
  present a short status of interferometry today. Then I detail the
  possible astrophysical prospects. Finally I address some important
  instrumental issues that are decisive for the future of
  interferometry.}
\motscles{Astronomie, interférométrie optique, instruments
  astronomique}

\begin{document}
\maketitlepage

\section{Introduction}

When I was asked to give an invited review on the topic of the
future of optical interferometry, I was tempted to give a short
answer. Everybody knows where to go (see contribution of A.\ Quirrenbach
in this volume) and there is no needs to detail
these directions of developments:
\begin{itemize}
\item \textbf{Higher spatial resolution} meaning going from
  milli-arcsecond scale to micro-arcsec ones.
\item \textbf{Higher flux sensitivity} meaning going beyond the
  Galaxy and reaching objects brighter than $K=13$.
\item \textbf{Higher astrophysical complexity} meaning going from
  \emph{visibilities} to \emph{true} images.
\end{itemize}
Therefore these advances require \textbf{many more} telescopes,
\textbf{much larger} apertures, \textbf{much longer} baselines in
excellent ground-based sites and eventually in space.

\begin{center}
  \emph{Will it be that easy?}
\end{center}

We have to remind us that radio interferometry took more than 30 years
from the first attempts in mid-1940's \cite{1960MNRAS.120..220R} to
the \emph{Very Large Array} in the mid-1970's
\cite{1980ApJS...44..151T}. Moreover, there is a $100\,000$ ratio
between the H$_{\rm I}$ wavelength at 21\,cm and the Bracket $\gamma$
line at $2.165\microns$ and therefore a similar ratio in accuracy
requirements. Interference detection in radio is done using the
heterodyne technique whereas in the optical domain one has to mix
first the optical signal before directly measuring it.

In addition, optical interferometry requires nanometer precision over
hundreds of meters, high reliability, complex instrumental control
using active control loops working at the kilo-hertz frequency, and
most of all, the atmosphere is corrugating the incoming wavefront on
centimeter scale on milli-second temporal scales.

Starting from the present state of the art (Sect.~\ref{sect:today}), I
present the directions where one can reasonably think that optical
interferometry can extend the parameter space in astrophysics
(Sect.~\ref{sect:astro}) depending on which instrumental issues
(Sect.~\ref{sect:instr}). 

\section{Where do we stand today?}
\label{sect:today}

On the hardware side, today astronomers using optical interferometry
have access to baselines ranging from 10 to 350\,m, aperture diameters
ranging from 50\,cm to 8-10\,m, detection wavelength ranging from 1 to
$10\,\microns$ (a few years ago even the window $0.4-0.8\,\microns$ was
accessible at SUSI and GI2T) and space interferometry vessels have not
yet been launched (SIM should be launched around 2015).   

The observables accessible to general astronomical users range from
squared amplitudes of the visibilities, color-differential phases,
closure phases to dual phases. These quantities allow the astronomers
to perform model fitting at different levels of complexity. Imaging
has been demonstrated but it is not really a routine procedure like
with radio interferometers. We are experiencing the premises of the
nulling technique and narrow-angle astrometry.

\begin{figure}[t]
  \centering
  \includegraphics[width=0.48\hsize]{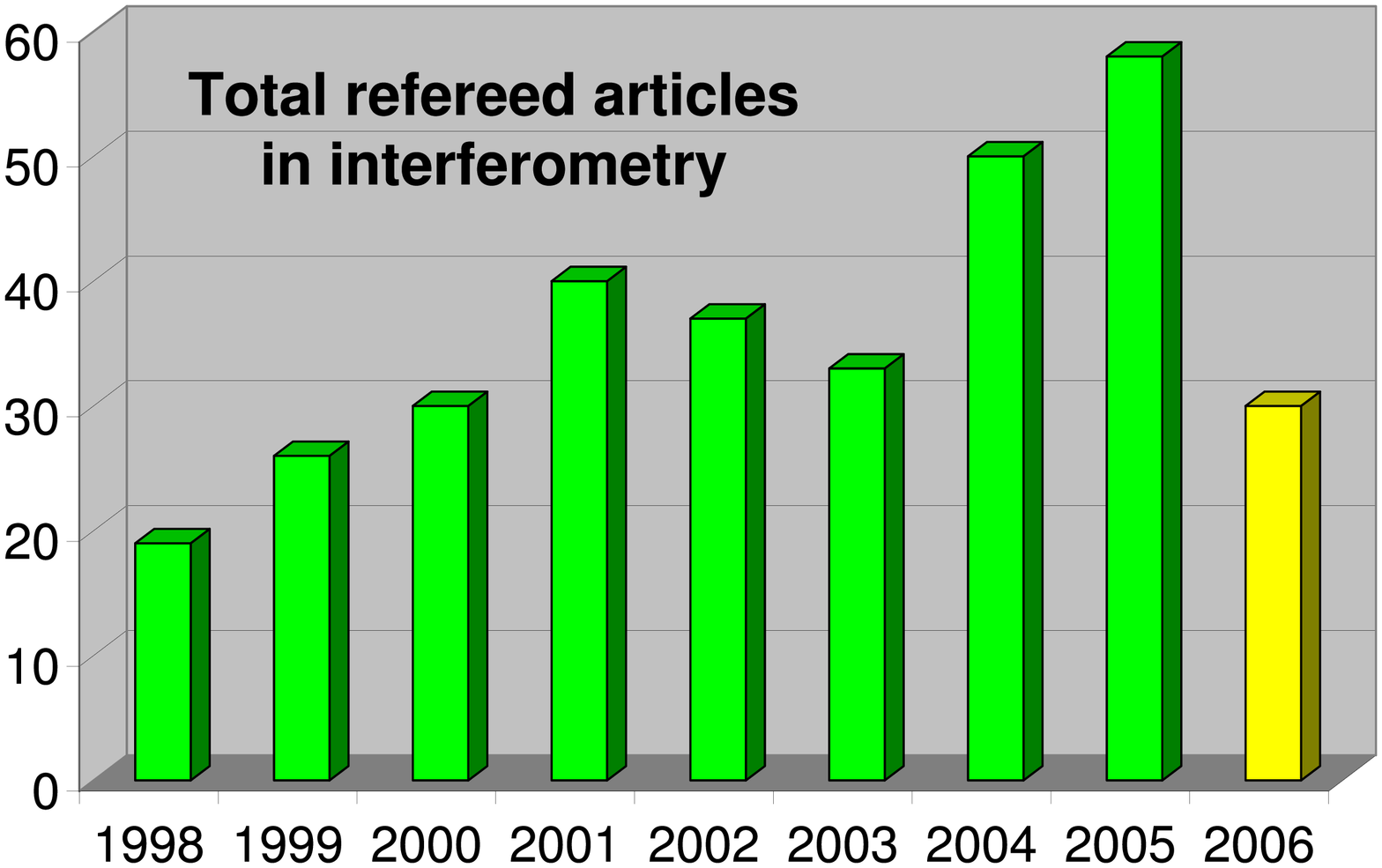}
  \hfill
  \includegraphics[width=0.48\hsize]{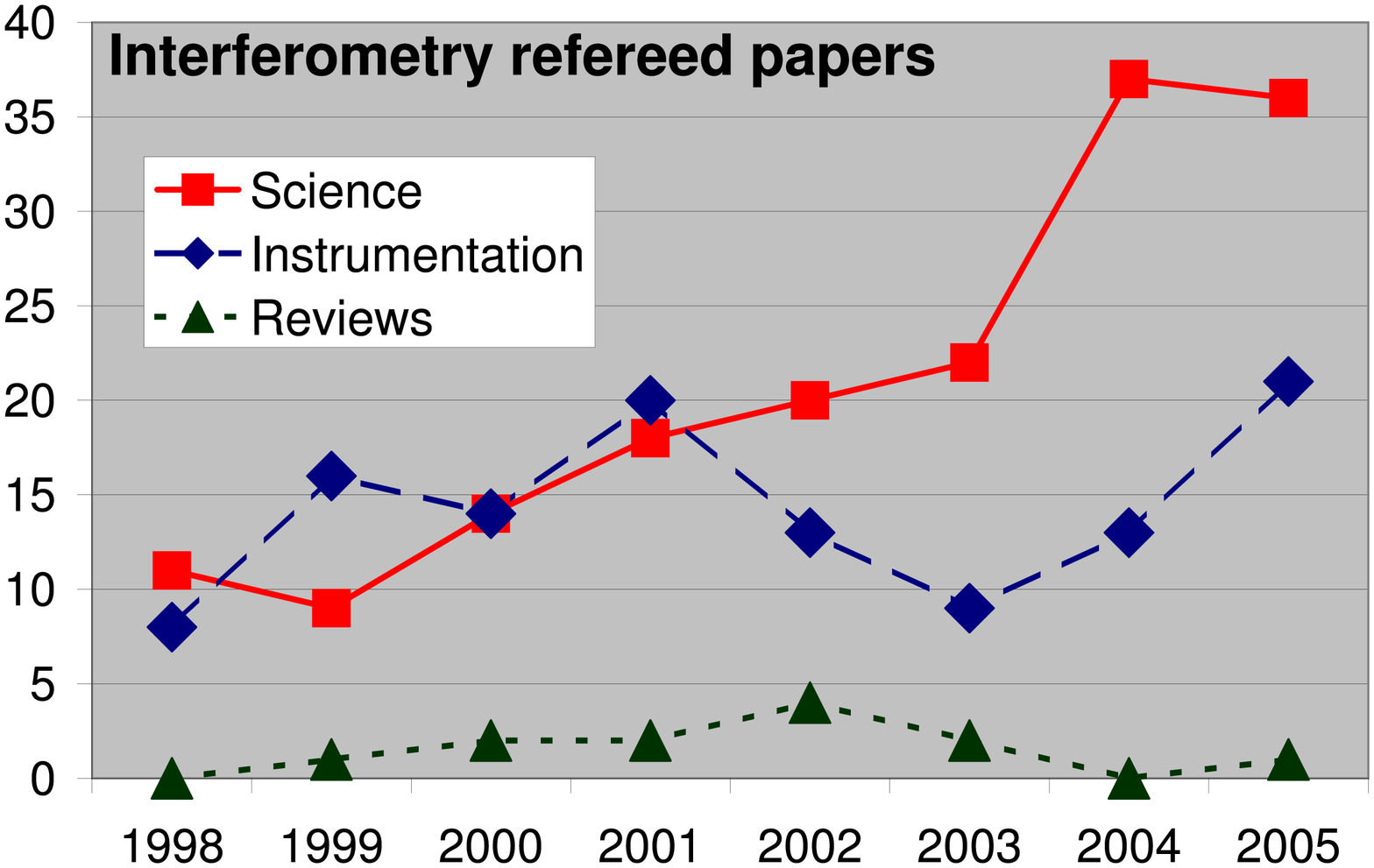}
  \caption{Refereed articles in optical interferometry (source:
    OLBIN). Left:
    evolution of number of refereed articles with years. Right:
    distribution of these papers between the different types of papers.} 
  \label{fig:refpapers}
\end{figure}

Even with a rather limited scope, we have been experiencing an giant leap in
the progress of optical interferometry demonstrated by the huge amount
of new astrophysical results obtained during the last years (see
graphs of Fig.~\ref{fig:refpapers} excerpted from
OLBIN\footnote{\texttt{http://olbin.jpl.nasa.gov}} database). 

Most of the astrophysical results in optical interferometry remains in
stellar physics: stellar diameters, circumstellar environments,
multiple systems,... However other fields are emerging like
extragalactic studies with the advent of large aperture
interferometers \cite{2003ApJ...596L.163S,2004Natur.429...47J}. With
increased accuracy, interferometers can measured stellar diameters of
even the lowest mass stars \cite{2003A&A...397L...5S}.

\section{Astrophysical prospects}
\label{sect:astro}

I do not detail here all the achievements obtained by
optical interferometers in the astrophysical field since S.\ Ridgway
in this volume already tackled this issue.

Although several orders of magnitude of spatial scales may reveal the
same physics, very strong changes can occur within a simple factor 10
in spatial resolution. A good example is the case of IRC 10216 which
has been observed at large scales by \citename{1999A&A...349..203M}
\citeyear{1999A&A...349..203M} with the \emph{Hubble Space
  Telescope} with a field of 2 arcminutes and at spatial resolution by
\citename{2002A&A...392..131W} \citeyear{2002A&A...392..131W} with
a field of 1 arcsecond and a resolution of 50-100 milli-arcseconds. At
large scales, the source appears centrosymmetric with spherical wind
whereas at high spatial resolution the source is clumpy and changes at
the year scale.

\begin{figure}[t]
  \centering
  \includegraphics[width=0.6\hsize]{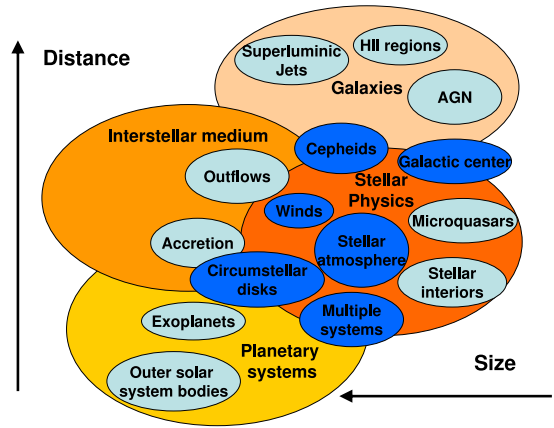}
  \caption{Parameter space for astrophysical prospects in optical
  interferometry. Dark colored items corresponds to fields already
  tackled by today interferometry. Light colored items are possible
  extension areas.}
  \label{fig:interf-fields}
\end{figure}

\subsection{Stellar physics}

Observations with optical interferometry goes already beyond the
measurement of stellar diameters with the observation of many classes
of stars (young, evolved, multiple, main-sequence, massive,
low-mass,...) and close phenomena like accretion, outflows, every sort
of shells. Some studies of the stellar surfaces have already been
achieved. 

In the near future, stellar atmosphere climatology, i.e.\ the study of dark
or hot spots, should be reachable and will provide new inputs to
Doppler imaging techniques. Convection in stellar interiors should be
within reach and will provide complementary evidences to those
obtained with asteroseismology techniques. 

Since the sensitivity of interferometers but also their performance
should improve, then an exploration of the whole Hertzsprung-Russel
diagram should be feasible from low-mass protostars to the remnants of
stellar evolution. Similarly, the connection with interstellar medium
can be contemplated by studying the influence of accretion of gas and
dust material as well as reversely the consequences of the ejection of
stellar matter.

In conclusion, we can imagine that the progress of optical
interferometry both in the visible and the infrared wavelength ranges
will open avenues to the understanding of the formation of stars and
planets as well as to the comprehension of the fate of stars.

\subsection{Planetary science}

The field of planetary science will undoubtedly benefit from the
progress in optical interferometry. The objects located at the
outskirts of the solar system like the Trans-Neptunian Objects (TNOs)
or those within the Kuiper belt are usually too small to be observed
by classical techniques. Adaptive optics have shown that some of them
can be spatially resolved from Earth, yet optical interferometry
should be able to improve the spatial resolution of these objects.

More than 10 years ago a new field has opened up: the study of planets
in extra-solar systems, also called exoplanets. Their proximity to
their stellar host and the contrast of their brightness with the one
of the central star make them difficult targets of
observation. However, observations of protoplanetary disks which are
probably the nurseries of such planets but also of the zodiacal light
of these planetary systems should be possible. The formation of
extrasolar systems is the occasion to observe the migration of giant
planets and the planetary gaps opened by the formation of rocky planets.

Observations could be enlarge to the different types of objects so
that we can explore the zoo of extrasolar planets. Interferometry
might be a tool to explore the extent of habitable zone around close
stars. Imaging of exo-Earth might seem a dream now but can be
contemplated using interferometry techniques. A mission like
DARWIN/TPF will certainly bring us clues on the status of extra
terrestrial life out side the Solar System.

\subsection{Interstellar medium}

The interstellar medium has been barely examined with optical
interferometry whereas it is one of the pillar of the science performed
by radio interferometry. However, prospects can be imagined that will
allow to observe the impacts of stellar outflows parsec away from the
central jet engine traced by shocks. This is also the location of wind
collision when two massive stars have strong mass loss.

By observing the dynamics of stellar clusters, either
dense ones or located far away, optical interferometry can bring
pieces of knowledge to the star formation and evolution. The increase
in spatial resolution will allow astronomers to observe H$-{\rm II}$
region in nearby galaxies.

\subsection{Extragalactic exploration}

As written above, optical interferometry has already started to
observe much more distant objects like the central cores of galaxies
like active galactic nuclei (AGN). In nearby galaxies, the brightest
objects should be within reach: Cepheids, supernovae, massive star
formation, but also fainter ones. It opens the possibility to study
galaxy dynamics. This already started with the observation of the
center of our Galaxy \cite{2005AN....326R.569P}.

Using the brightest extragalactic sources, interferometry should be able
to observe the details of superluminic jets being complementary to
radio Very Long Baseline Interferometry (VLBI). In addition with
spectral information, very high spatial observations of super-massive
black holes should also be possible. Finally one can imagine to extent
the interferometry techniques to the X-ray domain.

\subsection{Top level requirements}

The requirements for optical interferometry in the current fields of
astrophysics in order to increase the knowledge of these objects are
\textbf{access to routine imaging}, \textbf{observe in moderate and
  high spectral resolution}, \textbf{increasing the spectral range}
and \textbf{improve the dynamic range of observations}. To enlarge the
field of investigation of interferometry toward the interstellar
medium, \textbf{larger fields of view} and observations in
\textbf{mid-infrared to far infrared} are required. \textbf{Higher sensitivity}
and \textbf{longer baselines} are necessary to observe more distant objects
like galaxies and quasars, but also to observe with enhanced spatial
resolution the surface of stars. In the latter case but also in high
energy physics like environments of black-holes, \textbf{access to shorter
wavelengths} down to X-rays should be a priority.

\section{Instrumental issues}
\label{sect:instr}

I have listed above a list of astrophysical prospects and their
translation in top level requirements. But how do these requirements
have an implication into instrumental development? For example,
routine imaging is necessary because modeling cannot answer all
questions and imaging requires many telescopes, but how many of them?
Higher spatial resolution implies long baselines but how long, or
shorter wavelength but how short: ultraviolet, X-rays? 

Increasing the sensitivity is a key issue, but also the contrast
between the brightest object and the faintest one. Off-axis references
can be used, but requires special hardware like the PRIMA facility in
the VLTI. Certainly improving the capabilities a detector is also a
path to follow and since the sensitivity is more or less independent
of aperture size for ground-based interferometers, finding the best
site is crucial. Space-based instruments should also be continued to
be contemplated.  However, improving single pieces of hardware is not
sufficient and we must pay attention to the interferometer as a whole.

Optical interferometers are not really complicated (made of numerous
elements intricately combined) but remains complex (composed of several
interconnected units). For example the VLTI in the current state is made of:
\begin{itemize}
\item \textbf{Light collectors}: telescopes, guiding and active optics
\item \textbf{Beam routing optics}: 32 motors are involved
\item \textbf{Adaptive optics}: consisting in wavefront sensors,
  deformable mirrors, real-time controllers
\item \textbf{Delay lines}: 3 translation stages, metrology, switches,
  control to manage carriage trajectories
\item \textbf{Beam stabilizers}: variable curvature mirrors, image and
  pupil sensors (ARAL/IRIS), sources (LEONARDO)
\item \textbf{Fringe tracker}: fringe sensor, optical path difference
  controller managing fringe search, group delay or phase tracking
\item \textbf{Beam combination}: mainly the instruments VINCI, AMBER
  or MIDI which have a variety of spectral dispersion, spectral
  coverage, spatial filtering, detection and should control also the
  atmospheric dispersion and polarization
\item \textbf{Control software}: 60 computers and 750,000 lines of
  code \cite{2004SPIE.5496...21W}.
\end{itemize}
This is only for the combination of 2 (MIDI) or 3 (AMBER) beams!
Complexity follows the power of number of apertures and therefore
combining more telescopes like several tens will be a major challenge
and have a high price if one does not change the type of technology
used.  However with increasing number of telescopes, the impact of
failure is also less important than for small number of telescopes
especially if the setup is redundant enough.

Calibration is an important step of present interferometry. New
advances should take into account this step which becomes decisive in
a good implementation plan. For examples, interferometers with very
long baselines must be prepared to calibrate very low visibilities
using baseline bootstrapping or other methods. Another example are
spaced-based interferometers which absolutely needs to calibrate their
configuration before moving to another one since it is highly
improbable they can reconfigure exactly the same way. In these
perspective, spectral calibration may become an interesting way of
calibrating interferometric measurements on the science target itself.

In that respect, one determining question is how to combine several
tens of beams: using aperture synthesis like today or should we push
forward direct imaging? Aperture synthesis imaging requires less
telescopes at once, that can be compensated by more observing
time. This is probably the first step to imaging. Direct combination
imaging is simpler to manage and more photon efficient, but requires
at least a few tens of apertures and homothetic pupil combination
although densified pupil technique is possible.

Other parameters should be taken into account, like the
complementarity of diluted versus filled extremely large telescopes
(see contribution of Quirrenbach in this volume) which is an already
known story. In fact, in my opinion, this has been one the major achievements
of P.~Léna to succeed with his colleagues in convincing the European
astronomical community to build the \emph{Very Large Telescope} as an
interferometer. We should not forget also to continue to prepare the
path to space-based interferometers (see contributions of Fridlund and
Ollivier in this volume). 

\section{Conclusion}

We can conclude that there is certainly a future for optical
interferometry! However it will not be as straightforward as initially
and usually thought. While it is crucial to support the astrophysical
use of current interferometric facilities, it is also essential to
prepare the future (post-VLTI, KI, CHARA facility, ALOA), to identify
which is the most suited site, to continue developing interferometry
for space and of course to increase the number of apertures, the
baseline length, the size of the apertures to access to even larger
astrophysical topics.

\acknowledgements{I would like to warmly thanks Pierre Léna for his
  dedication and his commitments to the development of optical
  interferometry in Europe. On a more personal side, I would like to
  thank Pierre who have introduced me into interferometry and has been
  and still is an example to follow.}


\bibliography{malbet-vira}        

\end{document}